\documentclass[11pt]{article}
\usepackage[utf8]{inputenc}
\usepackage{amsfonts}
\usepackage{amsmath}
\usepackage{amssymb}
\usepackage{indentfirst}
\usepackage{graphicx}
\usepackage[colorlinks]{hyperref}
\usepackage{cite}
\usepackage{colortbl}
\usepackage{physics}
\usepackage{microtype}
\usepackage[normalem]{ulem}

\numberwithin{equation}{section}
\allowdisplaybreaks

\begin{document}

\baselineskip=18pt %a la harvmac
\baselineskip 0.6cm
\begin{titlepage}
\vskip 4cm

\begin{center}
\textbf{\LARGE Asymptotic structure of three-dimensional Maxwell Chern-Simons gravity coupled to spin-3 fields}
\par\end{center}{\LARGE \par}

\begin{center}
	\vspace{1cm}
    \textbf{Patrick Concha}$^{\dag, \star}$,
    \textbf{Javier Matulich}$^{\ast, \star}$,
    \textbf{Daniel Pino}$^{\ddag, \star}$,
	\textbf{Evelyn Rodríguez}$^{\dag, \star}$
	\small
	\\[5mm]
     $^{\dag}$\textit{Departamento de Matemática y Física Aplicadas, }\\
	\textit{ Universidad Católica de la Santísima Concepción, }\\
\textit{ Alonso de Ribera 2850, Concepción, Chile.}
  \\[2mm]
  $^{\star}$\textit{Grupo de Investigación en Física Teórica, GIFT, }\\
	\textit{Concepción, Chile.}
      \\[2mm]
	$^{\ast}$\textit{Instituto de Física Teórica UAM/CSIC,}\\
	\textit{C/ Nicolás Cabrera 13-15, Universidad Autónoma de Madrid, Cantoblanco, Madrid 28049,Spain. }
  \\[2mm]
	$^{\ddag}$\textit{Departamento de Física, Universidad de Concepción,}\\
	\textit{Casilla 160-C, Concepción, Chile.}
 \\[5mm]
	\footnotesize
	\texttt{patrick.concha@ucsc.cl},
    \texttt{javier.matulich@csic.es},
    \texttt{dpino2018@udec.cl},
	\texttt{erodriguez@ucsc.cl}

	\par\end{center}
\vskip 26pt
%\begin{abstract}
\centerline{{\bf Abstract}}
\medskip
\noindent  
In this work we analyze the asymptotic symmetries of the three-dimensional Chern-Simons (CS) gravity theory for a higher spin extension of the so-called Maxwell algebra. We propose a generalized set of asymptotic boundary conditions for the aforementioned flat gravity theory and we show that the corresponding charge algebra defines a higher-spin extension of the $\mathfrak{max}$-$\mathfrak{bms}_{3}$ algebra, which in turn corresponds the asymptotic symmetries of the Maxwell CS gravity. We also show that the $\mathfrak{hs}_3\mathfrak{max}$-$\mathfrak{bms}_{3}$ algebra can alternatively be obtained as a vanishing cosmological constant limit of three copies of the $\mathcal{W}_3$ algebra, with three independent central charges.

%\end{abstract}
\end{titlepage}\newpage {\baselineskip=12pt \tableofcontents{}}

\section{Introduction}\label{sec1}

The Maxwell symmetry, as well as its extensions with a non-vanishing cosmological constant, have received a growing interest in the context of (super)gravity during the last years. The Maxwell algebra can be defined in any spacetime dimension and can be obtained by considering an extension and deformation of the Poincaré symmetry. It was first introduced in the literature to describe the presence of a constant classical electromagnetic background in Minkowski spacetime \cite{Schrader:1972zd,Bacry:1970du,Gomis:2017cmt}.  In three spacetime dimensions, a gravity theory invariant under this symmetry can be constructed using the Chern-Simons (CS) formulation of gravity \cite{Salgado:2014jka,Hoseinzadeh:2014bla}. The CS action results to be given by three independent sectors, one of them given by the usual Einstein-Hilbert term without cosmological constant, while the other sectors are given by the exotic Lagrangian \cite{Witten:1988hc} and a term which involves the gravitational Maxwell field. The corresponding field equations are those of Poincaré gravity, describing a torsionless and flat spacetime, plus a third one involving the gravitational Maxwell gauge field. This theory has been deeply studied in different contexts such as (super)gravity theories \cite{Duval:2008tr,deAzcarraga:2010sw,Durka:2011nf,deAzcarraga:2012qj,Concha:2014xfa,Concha:2014tca,Penafiel:2017wfr,Ravera:2018vra,Concha:2018jxx,Concha:2018ywv,Concha:2019icz,Salgado-Rebolledo:2019kft,Chernyavsky:2020fqs,Kibaroglu:2020tbr,Cebecioglu:2021dqb}, higher-spin extensions \cite{Caroca:2017izc,Caroca:2021bjo,Caroca:2022byi}, non- and ultra-relativistic gravity theories \cite{Aviles:2018jzw,Concha:2019mxx,Gomis:2019nih,Concha:2020sjt,Concha:2020ebl,Concha:2021jos} and asymptotic symmetries \cite{Concha:2018zeb,Concha:2019eip,Adami:2020xkm}. 

Within the context of higher-spin (HS) gravity in three-dimensional spacetime, it was found in \cite{Caroca:2017izc} the extension with spin-3 gauge fields of Maxwell CS gravity. The underlying symmetry corresponds to a spin-3 extension of the Maxwell algebra, denoted here as $\mathfrak{hs_3}$$\mathfrak{max}$, allowing the inclusion of a new gauge field being the spin-3 analogue of the gravitational Maxwell field. Interestingly, the $\mathfrak{hs_3}$$\mathfrak{max}$ algebra can also be obtained as the vanishing cosmological constant limit of three copies of the $\mathfrak{sl}(3,\mathbb{R})$ algebra. Higher-spin fields have been of great interest due to its appearance in the spectrum of string theory and simplified models of the AdS/CFT conjecture \cite{Sundborg:2000wp,Klebanov:2002ja,Sezgin:2002rt,Sorokin:2004ie,Bekaert:2012ux,Gaberdiel:2012uj,Giombi:2012ms,Gaberdiel:2014cha,Rahman:2015pzl,Giombi:2016ejx}. In particular, in three-dimensions the coupling of massless HS fields to anti-de-Sitter (AdS) gravity is consistently described by a CS action whose gauge group is given, in the simplest case, by two copies of $SL(3,\mathbb{R})$ \cite{Blencowe:1988gj,Bergshoeff:1989ns,Henneaux:2010xg}, which describes nonpropagating spin-3 fields coupled to AdS gravity. Despite the lack of local degrees of freedom, CS theories exhibit a rich structure that merits further investigation. In fact, akin to the situation in pure gravity, the $SL(3,\mathbb{R}) \times SL(3,\mathbb{R})$ CS theory features interesting solutions, including HS black holes \cite{Gutperle:2011kf,Perez:2012cf,Perez:2013xi,Compere:2013gja,Bunster:2014mua,Banados:2015tft,Grumiller:2016kcp,Banados:2016hze} and conical singularities \cite{Castro:2016ehj,Campoleoni:2013iha} . As explained in \cite{Matulich:2014hea} it is possible to perform the vanishing cosmological constant limit in a straightforward way when an appropriate gauge is chosen, so that the HS black hole solution reduces to a HS extension of locally flat cosmological spacetimes \cite{Barnich:2012xq,Bagchi:2012xr}. Furthermore, the asymptotic symmetry of the HS AdS theory realizes two copies of the centrally-extended $\mathcal{W}_3$ algebra \cite{Henneaux:2010xg, Campoleoni:2010zq }, whose flat limit was studied  in \cite{Afshar:2013vka,Gonzalez:2013oaa,Gary:2014ppa,Matulich:2014hea}. 

One of the recent results showed that the gravitational Maxwell field modifies not only the vacuum of the CS theory but also its asymptotic structure \cite{Concha:2018zeb}. The asymptotic symmetry algebra in this case is described by a deformed $\mathfrak{bms}_{3}$ algebra, denoted in this work as $\mathfrak{max}$-$\mathfrak{bms}_{3}$, and was first introduced in \cite{Caroca:2017onr} as an $S$-expansion of the Virasoro algebra.  Asymptotic symmetries are key physical symmetries in the theory that significantly influence the state of the system. Its relevance becomes even more important in topological theories like three-dimensional gravity, where the dynamic is entirely captured by boundary degrees of freedom and holonomies. Therefore, grasping the asymptotic dynamics of extended (super)gravity theories is crucial, for instance, for developing dual theories for three-dimensional extended supergravities. This topic is especially compelling, as it could provide valuable insights into holography in non-AdS or Poincaré contexts.

The purpose of this work is to extend the asymptotic conditions considered for Maxwell CS gravity in \cite{Concha:2018zeb}, to include spin-3 gauge fields non-minimally coupled to the theory. Then, our results can be seen as a novel set of asymptotic boundary conditions for higher spin gravity without cosmological constant in three dimensions. 
The paper is organized as follows: In section 2, we briefly review the three-dimensional HS extension of the Maxwell CS gravity, which is based on the $\mathfrak{hs_3}$$\mathfrak{max}$ symmetry. In section 3, we provide the asymptotic symmetry algebra for the minimal HS Maxwell CS gravity, which results to be given by a HS extension of the deformed $\mathfrak{bms}_3$ algebra, with three central charges.  We propose a suitable fall-off conditions for the gauge fields at infinity and the gauge transformations preserving the boundary conditions. In section 4, we explicitly show that the $\mathfrak{hs_3}$$\mathfrak{max}$-$\mathfrak{bms}_3$ algebra can be found as the vanishing cosmological constant limit $\ell\rightarrow\infty$ of three copies of the $W_3$ algebra.  Section 5 is devoted to some discussion and possible future developments. 

\section{Three-dimensional Maxwell Chern-Simons gravity coupled to spin-3 fields}
In this section, we briefly review the three-dimensional Maxwell CS gravity coupled to spin-3 fields first presented in \cite{Caroca:2017izc}. The CS gravity action is constructed from the spin-3 extension of the Maxwell algebra, referred in this work as $\mathfrak{hs_{3}max}$, whose generators satisfy the following non-vanishing commutators:
   \begin{alignat}{2}
        [J_a,J_b] &= \epsilon_{abc} J^{c},\qquad \qquad & [J_a,P_b] &= \epsilon_{abc} P^{c},\, \notag\\
        [P_a,P_b] &= \epsilon_{abc}Z^{c},\qquad \qquad &[J_a,Z_b] &= \epsilon_{abc}Z^{c},\, \notag\\
        [J_a,J_{bc}] &= \epsilon_{a(b}^{m}J_{c)m}, \qquad \qquad  & [J_a,P_{bc}] &= \epsilon_{a(b}^{m} P_{c)m},\, \notag\\
        [P_a,J_{bc}] &= \epsilon_{a(b}^{m}P_{c)m}, \qquad \qquad & [P_a,P_{bc}] &= \epsilon_{a(b}^{m}Z_{c)m},\, \notag\\
        [Z_{a},J_{bc}] &= \epsilon_{a(b}^{m}Z_{c)m}, \qquad \qquad & [J_a,Z_{bc}]& = \epsilon_{a(b}^{m}Z_{c)m},\, \notag\\
        [J_{ab},J_{cd}] &= - \qty(\eta_{a(c}\epsilon_{d)bm} + \eta_{b(c}\epsilon_{d)am})J^{m},\, \notag\\
        [J_{ab},P_{cd}] &= - \qty(\eta_{a(c}\epsilon_{d)bm} + \eta_{b(c}\epsilon_{d)am})P^{m},\, \notag\\
        [J_{ab},Z_{cd}] &= - \qty(\eta_{a(c}\epsilon_{d)bm} + \eta_{b(c}\epsilon_{d)am}) Z^{m},\, \notag\\
        [P_{ab},P_{cd}] &= - \qty(\eta_{a(c}\epsilon_{d)bm} + \eta_{b(c}\epsilon_{d)am}) Z^{m},\,
\end{alignat}
where $a,b,\dots =0,1,2$ are Lorentz indices raised and lowered with the  Minkowski metric $\eta _{ab}$ and $\epsilon _{abc}$ is the Levi-Civita tensor. This algebra turns out to be a generalization of the Maxwell algebra and describes the coupling of spin-3 generators $\lbrace J_{ab},P_{ab},Z_{ab}\rbrace$ to the Maxwell ones $\lbrace J_{a},P_{a},Z_{a}\rbrace$. It was first derived as an expansion \cite{Izaurieta:2006zz} of the $\mathfrak{sl}(3,\mathbb{R})$ algebra with a particular semigroup, and is naturally recovered through an Inönü-Wigner contraction of three copies of $\mathfrak{sl}(3,\mathbb{R})$. It is important to recall that the spin-3 generators are assumed to be symmetric and traceless.

In order to write down a CS action for this algebra, we define the one-form gauge connection
\begin{equation}
A=\omega^{a}J_a+e^{a}P_{a}+\sigma^{a}Z_{a}+\omega^{ab}J_{ab}+e^{ab}P_{ab}+\sigma^{ab}Z_{ab}\,,\label{oneformspin3Max}
\end{equation}
where $e^{ab},\omega^{ab}$ and $\sigma^{ab}$ correspond to the spin-3 analogues of the vielbein, spin connection and Maxwell field, respectively. The non-vanishing components of the invariant tensor are given by:
\begin{align}
\left\langle J_{a}J_{b}\right\rangle  &=\alpha _{0}\eta _{ab}\,, &
\left\langle P_{a}P_{b}\right\rangle  &=\alpha _{2}\eta _{ab}\,, \notag \\
\left\langle J_{a}P_{b}\right\rangle  &=\alpha _{1}\eta _{ab}\,, &
\left\langle J_{a}Z_{b}\right\rangle  &=\alpha _{2}\eta _{ab}\,, \label{it1}
\end{align}
\begin{align}
\left\langle J_{ab}J_{cd}\right\rangle  &=\alpha _{0}\left( \eta _{a(c}\eta _{d)b}-\frac{2}{3}\eta
_{ab}\eta _{dc}\right) \,, &
\left\langle P_{ab}P_{cd}\right\rangle  &=\alpha _{2}\left( \eta _{a(c}\eta _{d)b}-\frac{2}{3}\eta
_{ab}\eta _{dc}\right) \,, \notag\\
\left\langle J_{ab}P_{cd}\right\rangle  &=\alpha _{1}\left( \eta _{a(c}\eta _{d)b}-\frac{2}{3}\eta
_{ab}\eta _{dc}\right) \,, &
\left\langle J_{ab}Z_{cd}\right\rangle  &=\alpha _{2}\left( \eta _{a(c}\eta _{d)b}-\frac{2}{3}\eta
_{ab}\eta _{dc}\right) \,.\label{it2}
\end{align}
Then, considering the previous invariant tensor and the one-form gauge 
connection (\ref{oneformspin3Max}) in the CS action 
\begin{equation}
I[A]=\frac{k}{4\pi }\int_{\mathcal{M}}\left\langle AdA+\frac{2}{3}
A^{3}\right\rangle \,,  \label{CSaction}
\end{equation}
defined on a three-dimensional manifold $\mathcal{M}$, and where $k=\frac{1}{4G}$ is the level of the theory related to the gravitational constant $G$, we obtain
\begin{eqnarray}
I_{\mathfrak{hs_{3}Max}} &=&\frac{k}{4\pi}\int \alpha _{0}\left[ \left( \omega ^{a}d\omega
_{a}+\frac{1}{3}\epsilon _{abc}\omega ^{a}\omega ^{b}\omega ^{c}\right)
+2 \left( \omega _{\text{ }b}^{a}d\omega _{\text{ }a}^{b}+2\epsilon
_{abc}\omega ^{a}\omega ^{bd}\omega _{\text{ }d}^{c}\right) \right]   \notag
\\
&&+2\alpha _{1}\left[ e^{a}\left( d\omega _{a}+\frac{1}{2}\epsilon _{abc}\omega
^{b}\omega ^{c}+2 \epsilon _{abc}\omega ^{bd}\omega _{\text{ }%
d}^{c}\right) +2 e^{ab}\left( d\omega _{ab}+2\epsilon _{acd}\omega
^{c}\omega _{\ b}^{d}\right) \right]   \notag \\
&&+\alpha _{2}\left[ e^{a}\left( de_{a}+\epsilon _{abc}\omega
^{b}e^{c}\right) +2\sigma^{a}\left( d\omega _{a}+\frac{1}{2}\epsilon _{abc}\omega
^{b}\omega ^{c}\right) +2 e_{\text{ }}^{ab}\left( de_{ab}+2\epsilon
_{acd}\omega ^{c}e_{\text{ }b}^{d}+4\epsilon _{acd}e^{c}\omega _{\text{ }%
b}^{d}\right) \right.   \notag \\
&&\left. +4\left( \omega ^{ab}d\sigma_{ab}+\epsilon _{abc}\sigma^{a}\omega
^{be}\omega _{\text{ }e}^{c}+2\epsilon _{abc}\omega ^{a}\sigma^{be}\omega _{\text{
}e}^{c}\right) \right] \,.  \label{Maxspin3}
\end{eqnarray}
This CS action describes the coupling of spin-3 gauge fields to three-dimensional Maxwell gravity and corresponds to a novel extension of higher-spin three-dimensional gravity in flat space including topological HS matter. It has three different independent sectors
proportional to $\alpha _{0},\alpha _{1}$ and $\alpha _{2}$. The
term proportional to $\alpha _{1}$ corresponds to an Euler type CS form while the term proportional to $\alpha _{0}$ and $\alpha _{2}$ are Pontryagin type CS forms. As it was mentioned in \cite{Caroca:2017izc}, similarly to the spin-3 extension of the Poincar\'{e} gravity, the action (\ref{Maxspin3}) does
not contain the cosmological constant term. Extremization of the action gives rise to the following field equations for the
spin-2 fields
\begin{eqnarray}
\mathcal{T}^{a} &\equiv &de^{a}+\epsilon ^{abc}\omega _{b}e_{c}+4
\epsilon ^{abc}e^{bd}\omega _{c}^{\text{ }d}=0\,,\notag \\ 
\mathcal{R}^{a} &\equiv &d\omega ^{a}+\frac{1}{2}\epsilon ^{abc}\omega
_{b}\omega _{c}+2 \epsilon ^{abc}\omega _{bd}\omega _{c}^{\text{ }
d}=0\,,\notag \\ 
\mathcal{F}^{a} &\equiv &d\sigma^{a}+\epsilon ^{abc}\omega _{b}\sigma_{c}+\frac{1}{2}%
\epsilon ^{abc}e_{b}e_{c}+2 \epsilon ^{abc}\left( 2\omega _{bd}\sigma_{c}^{%
\text{ }d}+e_{bd}e_{c}^{\text{ }d}\right) =0\,,\label{hsm1}
\end{eqnarray}%
while the corresponding field equations for the spin-3 fields are
\begin{eqnarray}
\mathcal{T}^{ab} &\equiv&de^{ab}+\epsilon ^{cd\left( a\right\vert }\omega
_{c}e_{d}^{\text{ }\left\vert b\right) }+\epsilon ^{cd\left( a\right\vert
}e_{c}\omega _{d}^{\text{ }\left\vert b\right) }=0\,, \notag \\ 
\mathcal{R}^{ab} &\equiv&d\omega ^{ab}+\epsilon ^{cd\left( a\right\vert }\omega
_{c}\omega _{d}^{\text{ }\left\vert b\right) }=0\,, \notag \\ 
\mathcal{F}^{ab} &\equiv&d\sigma^{ab}+\epsilon ^{cd\left( a\right\vert }\omega
_{c}\sigma_{d}^{\text{ }\left\vert b\right) }+\epsilon ^{cd\left( a\right\vert
}\sigma_{c}\omega _{d}^{\text{ }\left\vert b\right) }+\epsilon ^{cd\left(
a\right\vert }e_{c}e_{d}^{\text{ }\left\vert b\right) }=0\,.\label{hsm3}
\end{eqnarray} 
The CS action (\ref{Maxspin3}) is invariant, up to boundary terms, under the action of the infinitesimal gauge
transformations $\delta A=D\lambda=d\lambda+[A,\lambda]$, where the local gauge parameter is given by
\begin{equation}
\lambda =\Lambda ^{a}J_{a}+\xi ^{a}P_{a}+\chi ^{a}Z_{a}+\Lambda
^{ab}J_{ab}+\xi ^{ab}P_{ab}\,+\chi ^{ab}Z_{ab}\,.
\end{equation}
For the spin-2 gauge fields we get
\begin{eqnarray}
\delta \omega ^{a} &=&D_{\omega }\Lambda ^{a}\,+4 \epsilon
^{abc}\omega _{bd}\Lambda _{c}^{\text{ }d}\,, \notag \\
\delta e^{a} &=&D_{\omega }\xi ^{a}-\epsilon ^{abc}\Lambda _{b}e_{c}+4
\epsilon ^{abc}\omega _{bd}\xi _{c}^{\text{ }d}+4 \epsilon
^{abc}e_{bd}\Lambda _{c}^{\text{ }d}\,\,, \notag \\
\delta \sigma^{a} &=&D_{\omega }\chi ^{a}-\epsilon ^{abc}\xi _{b}e_{c}-\epsilon
^{abc}\Lambda _{b}\sigma_{c}+4 \epsilon ^{abc}e_{bd}\xi _{c}^{\text{ }d}
+4 \epsilon ^{abc}\omega _{bd}\chi _{c}^{\text{ }d}+4 \epsilon
^{abc}\sigma_{bd}\Lambda _{c}^{\text{ }d}\,,
\end{eqnarray}%
where, besides the usual gauge transformations of CS Maxwell gravity, there are new terms
involving the spin-3 gauge parameters $\xi ^{ab}$, $\Lambda ^{ab}$ and
$\chi ^{ab}$. On the other hand, the spin-3 gauge fields transform as follows
\begin{eqnarray}
\delta \omega ^{ab} &=&d\Lambda ^{ab}\,+\epsilon ^{cd\left( a\right\vert
}\omega _{c}\Lambda _{d}^{\text{ }\left\vert b\right) }+\epsilon
^{cd(a}\omega _{\text{ }c}^{b)}\Lambda _{d}\,, \notag \\
\delta e^{ab} &=&d\xi ^{ab}+\epsilon ^{cd\left( a\right\vert }\omega _{c}\xi
_{d}^{\text{ }\left\vert b\right) }+\epsilon ^{cd\left( a\right\vert
}e_{c}\Lambda _{d}^{\text{ }\left\vert b\right) }+\epsilon ^{cd(a}e_{\text{ }%
c}^{b)}\Lambda _{d}+\epsilon ^{cd(a}\omega _{\text{ }c}^{b)}\xi _{d}\,,\notag \\
\delta \sigma^{ab} &=&d\chi ^{ab}+\epsilon ^{cd\left( a\right\vert }\sigma_{c}\Lambda
_{d}^{\text{ }\left\vert b\right) }+\epsilon ^{cd\left( a\right\vert }\omega
_{c}\chi _{d}^{\text{ }\left\vert b\right) }+\epsilon ^{cd\left(
a\right\vert }e_{c}\xi _{d}^{\text{ }\left\vert b\right) }  \notag \\
&&+\,\epsilon ^{cd(a}\omega _{\text{ }c}^{b)}\chi _{d}+\epsilon ^{cd(a}\sigma_{%
\text{ }c}^{b)}\Lambda _{d}+\epsilon ^{cd(a}e_{\text{ }c}^{b)}\xi _{d}\,.
\end{eqnarray}

A consistent set of boundary conditions for the pure Maxwell gravity theory was proposed in \cite{Concha:2018zeb}, whose asymptotic symmetry algebra was shown to be given by a deformation of the $\mathfrak{bms}_3$ algebra, referred to here as $\mathfrak{max}$-$\mathfrak{bms}_3$, with three independent central charges. Subsequently, in \cite{Concha:2018jjj} it was shown that the $\mathfrak{max}$-$\mathfrak{bms}_3$ can be recovered from three copies of the Virasoro algebra, which in turn were shown to correspond to the asymptotic symmetry algebra of the AdS-Lorentz algebra.  In the next sections we shall extend these results to the previously discussed Maxwell CS gravity coupled with spin-3 gauge fields. We will include chemical potentials without spoiling the original deformed $\mathfrak{bms}_{3}$ symmetry. As we will see, the corresponding asymptotic symmetry algebra will correspond to a spin-3 extension of $\mathfrak{max}$-$\mathfrak{bms}_{3}$. The charge algebra has three central charges defined in terms of the coupling constants appearing in the CS action \eqref{Maxspin3}. We will also show that this asymptotic symmetry can alternatively be obtained through a well-defined flat limit of three copies of the $\mathcal{W}_3$ algebra.

\section{Asymptotic symmetry algebra }
In this section, following the methodology used in \cite{Matulich:2014hea} we derive the asymptotic symmetry algebra for the spin-3 Maxwell CS gravity theory. First, inspired by the boundary conditions we have to impose in the pure Maxwell gravity, we provide the suitable fall-off conditions for the gauge fields at infinity and the gauge transformations preserving our proposed boundary
conditions. Then, using the Regge-Teitelboim method \cite{Regge:1974zd} we find the charge algebra which, as it is expected, will be a higher spin  extension of the $\mathfrak{max}$-$\mathfrak{bms}_{3}$ algebra \cite{Caroca:2017onr,Concha:2018zeb}. The Poisson algebra structure we find is then obtained as a flat limit from a classical centrally extended $\mathcal{W}_3\oplus\mathcal{W}_3\oplus \mathcal{W}_3$ algebra.

\subsection{On the boundary conditions}
To start with, we propose the following behavior of the gauge fields at the boundary
\begin{equation}\label{AsMF}
A=h^{-1}dh+h^{-1}ah\,,
\end{equation}
where the radial dependence is entirely captured by the group element $h=$ e$^{-rP_{0}}$. The auxiliary gauge field $a$ has the form 
\begin{equation}
   a=a_{\phi}d{\phi} +a_udu\,,
\end{equation}
with the angular component given by
\begin{eqnarray}
a_\phi &= &  J_{1}+\frac{1}{2}{\cal N} P_{0}+\frac{1}{2}{\cal  M}\, J_{0}+\frac{1}{2}\mathcal{F}\,Z_{0} +{\cal V} P_{00}+{\cal  W}\, J_{00}+\mathcal{X}\,Z_{00}\,,\label{aa}
\end{eqnarray}
where the functions ${\cal M}$, ${\cal N}$, $\cal{F}$, $\cal{V}$, $\cal{W}$ and $\cal{X}$ are assumed to depend on all boundary coordinates $x^{i}=(u,\phi)$. Let us note that the form of the angular component generalizes the asymptotic behavior of the gauge fields considered in \cite{Gonzalez:2013oaa,Matulich:2014hea} to the presence of the gravitational Maxwell field along its spin-3 counterpart.

As it is well-known, the asymptotic symmetries correspond to the set of gauge transformations $\delta A = d\lambda + [A,\lambda]$
that preserve the asymptotic conditions (\ref{AsMF}), with
\begin{equation}
\lambda =\Lambda ^{a}(u,\phi)J_{a}+\xi ^{a}(u,\phi)P_{a}+\chi ^{a}(u,\phi)Z_{a}+\Lambda
^{ab}(u,\phi)J_{ab}+\xi ^{ab}(u,\phi)P_{ab}\,+\chi ^{ab}(u,\phi)Z_{ab}\,.
\end{equation}
Thus, the angular component $a_\phi$ is left invariant for the Lie-algebra-valued parameter $\lambda=\lambda(y,f,h,v,w,g)$ of the form
\begin{equation}\label{lambda}
    \begin{split}
    \lambda &= \qty(\frac{\mathcal{M}}{2}f+\frac{\mathcal{N}}{2}y + 4\mathcal{V}v + 4\mathcal{W}w-f'')P_0+ f P_1 -f'P_2+\qty( \frac{\mathcal{M}}{2}y + 4\mathcal{W}v-y'') J_0+ y J_1-y'J_2\\
    %----------------------------------------------------------------------------------
    &\quad + \qty(\frac{1}{2}\mathcal{M}h+\frac{1}{2}\mathcal{F}y+\frac{1}{2}\mathcal{N}f+4\mathcal{V}g+4\mathcal{X}w+4\mathcal{V}w-h'')Z_0+ h Z_1-h'Z_2\\
    %----------------------------------------------------------------------------------
    &\quad + \left[\frac{1}{6}w^{(4)} -\frac{2}{3}\mathcal{M}w''-\frac{7}{12}\mathcal{M}'w'+\frac{1}{12}(3\mathcal{M}^2-2\mathcal{M}'')w \right.\\
    &\qquad\left. -\frac{2}{3}\mathcal{N}v''-\frac{7}{12}\mathcal{N'}v'+
        \frac{1}{12}\qty(6\mathcal{MN}-2\mathcal{N}'')v+\mathcal{V}y+\mathcal{W}f \right]P_{00}\\
    &\quad +\qty(\mathcal{M}w+\mathcal{N}v-w'')P_{01}+\qty(\frac{1}{3}w'''  - \frac{5}{6}\mathcal{M}w' - \frac{1}{3}\mathcal{M}'w  - \frac{5}{6}\mathcal{N}v' - \frac{1}{3}\mathcal{N'}v) P_{02}\\
    &\quad +w P_{11}-w'' P_{12}+\qty[ \frac{v^{(4)}}{6}-\frac{2}{3}\mathcal{M}v''-\frac{7}{12}\mathcal{M'}v' +\frac{1}{12} \qty(3\mathcal{M}^2-2\mathcal{M}'')v+\mathcal{W}y] J_{00}\\
    &\quad+ \qty(
    \mathcal{M}v-v'') J_{01}+ \qty(\frac{v'''}{3} - \frac{5}{6}\mathcal{M}v'- \frac{1}{3}\mathcal{M}'v) J_{02}+ vJ_{11} -v' J_{12}\\
    %----------------------------------------------------------------------------------
    &\quad +\left[\frac{1}{6}g^{(4)}-\frac{2}{3}\mathcal{M}g''-\frac{7}{12}\mathcal{M}'g'+\frac{1}{12}\qty(3\mathcal{M}^2-2\mathcal{M}'')g+\mathcal{V}f+\mathcal{W}h+\mathcal{X}y-\frac{2}{3}\mathcal{F}v''-\frac{7}{12}\mathcal{F}'v' \right.\\
    &\qquad \left.+\frac{1}{12}\qty(6\mathcal{FM}+3\mathcal{N}^2-2\mathcal{F}'')v-\frac{2}{3}\mathcal{N}w''-\frac{7}{12}\mathcal{N}'w'+\frac{1}{12}\qty(6\mathcal{MN}-2\mathcal{N}'')w\right] Z_{00}\\
    &\quad +\qty(\mathcal{M}g+\mathcal{F}v+\mathcal{N}w-g'')Z_{01}+ \qty(\frac{1}{3}g'''-\frac{5}{6}\mathcal{M}g'-\frac{1}{3}\mathcal{M}'g-\frac{5}{6}\mathcal{F}v'-\frac{1}{3}\mathcal{F}'v-\frac{5}{6}\mathcal{N}w'-\frac{1}{3}\mathcal{N}'w)Z_{02}\\
    &\quad+ g Z_{11}-g'Z_{12}\,,
\end{split}
\end{equation}
provided the functions $\mathcal{M}, \mathcal{N}, \mathcal{F}, \mathcal{V}, \mathcal{W}, \mathcal{X}$ transform as follows
\begin{align}
    \delta\mathcal{M} &= \mathcal{M}'y +2\mathcal{M}y'-2y'''+4(2\mathcal{W}'v+3\mathcal{W}v')\,, \notag\\
    %-------------------------------------------------------------------------
    \delta \mathcal{N} &= \mathcal{M}'f+2\mathcal{M}f'+ \mathcal{N}'y + 2\mathcal{N}y' - 2f''' +4(2\mathcal{W}'w+3\mathcal{W}w')+ 4(2\mathcal{V}'v+3\mathcal{V}v')\,,\notag\\
   %-------------------------------------------------------------------------
    \delta\mathcal{F} &= \mathcal{M}'h+2\mathcal{M}h'+\mathcal{N}'f+2\mathcal{N}f'+\mathcal{F}'y+2\mathcal{F}y'-2h'''+4(2\mathcal{W}'g+3 \mathcal{W}g')\notag\\
    &\quad +4(2\mathcal{V}'w+3 \mathcal{V}w')+4(2\mathcal{X}'v+3 \mathcal{X}v'),\notag
    \\
    %-------------------------------------------------------------------------
    \delta\mathcal{W} &= \mathcal{W}'y + 3\mathcal{W}y' + \frac{1}{12}\qty(8\mathcal{M}\mathcal{M}' - 2\mathcal{M}''')v + \frac{1}{12}\qty(8\mathcal{M}^2-9\mathcal{M}'')v' -\frac{5}{4}\mathcal{M}'v'' - \frac{5}{6}\mathcal{M}v'''+ \frac{1}{6}v^{(5)}\,,\notag\\
    %-------------------------------------------------------------------------
    \delta\mathcal{V} &=\mathcal{W'}f+3\mathcal{W}f' +3\mathcal{V}y' +\frac{1}{12}(8\mathcal{NM'}+8\mathcal{MN'}-2\mathcal{N}''')v+\frac{1}{12}(16\mathcal{MN}-9\mathcal{N''})v'\notag\\
    &-\frac{5}{4}\mathcal{N}'v''-\frac{5}{6}\mathcal{N}v'''+\frac{1}{12}(8\mathcal{MM'}-2\mathcal{M}''')w+\frac{1}{12}(8\mathcal{M}^2-9\mathcal{M}'')w'-\frac{5}{4}\mathcal{M}'w''-\frac{5}{6}\mathcal{M}w'''+ \frac{1}{6}w^{(5)}\notag\\
    %-------------------------------------------------------------------------
    \delta\mathcal{X} &=\mathcal{W'}h+3\mathcal{W}h'+\mathcal{V}'f+3\mathcal{V}f'+\mathcal{X'}y+3\mathcal{X}y'+\frac{1}{12}\qty(8\mathcal{MF'}+8\mathcal{FM'}+8\mathcal{NN'}-2\mathcal{F}''')v\notag\\
    &+\frac{1}{12}\qty(16\mathcal{FM}+8\mathcal{N}^2-9\mathcal{F}'')v'-\frac{5}{4}\mathcal{F'}v''-\frac{5}{6}\mathcal{F}v'''+\frac{1}{12}\qty(8\mathcal{NM'}+8\mathcal{MN'}-2\mathcal{N}''')w\notag\\
    &+\frac{1}{12}\qty(16\mathcal{MN}-9\mathcal{N}'')w'-\frac{5}{4}\mathcal{N}'w''-\frac{5}{6}\mathcal{N}w'''\notag\\
    &+\frac{1}{12}\qty(8\mathcal{MM'}-2\mathcal{M}''')g+\frac{1}{12}\qty(8\mathcal{M}^2-9\mathcal{M}'')g'-\frac{5}{4}\mathcal{M}'g''-\frac{5}{6}\mathcal{M}g'''+\frac{1}{6}g^{(5)}\,,  
 \label{translaw}
\end{align}
where prime denotes the derivative with respect to the $\phi$ coordinate. 
The asymptotic symmetries along time will be preserved whenever the Lagrange multiplier is $A_{u}=h^{-1}a_{u}h$, with 

\begin{equation}\label{AsMu}
a_u=\lambda[ \mu,\xi, \vartheta,\varrho,\varepsilon,\varphi]\,,
\end{equation}
where $\mu,\xi, \vartheta,\varrho,\varepsilon$,  and $\varphi$ are arbitrary functions of $(u,\phi)$ which are assumed to be fixed at the boundary \cite{Henneaux:2013dra, Bunster:2014mua}. 
 The time  evolution of the gauge fields in the asymptotic region is given by the following conditions
\begin{align}
    \dot{\mathcal{M}} &= \mathcal{M}'\mu +2\mathcal{M}\mu'-2\mu'''+4(2\mathcal{W}'\varrho+3\mathcal{W}\varrho')\,, \notag\\
    %-------------------------------------------------------------------------
    \dot {\mathcal{N}} &= \mathcal{M}'\xi+2\mathcal{M}\xi'+ \mathcal{N}'\mu + 2\mathcal{N}\mu' - 2\xi''' +4(2\mathcal{W}'w+3\mathcal{W}w')+ 4(2\mathcal{V}'\varrho+3\mathcal{V}\varrho')\,,\notag\\
   %-------------------------------------------------------------------------
    \dot{\mathcal{F}} &= \mathcal{M}'\vartheta+2\mathcal{M}\vartheta'+\mathcal{N}'\xi+2\mathcal{N}\xi'+\mathcal{F}'\mu+2\mathcal{F}\mu'-2\vartheta'''+4(2\mathcal{W}'\varphi+3 \mathcal{W}\varphi')\notag\\
    &\quad +4(2\mathcal{V}'\varepsilon+3 \mathcal{V}\varepsilon')+4(2\mathcal{X}'\varrho+3 \mathcal{X}\varrho'),\notag
    \\
    %-------------------------------------------------------------------------
    \dot{\mathcal{W} }&= \mathcal{W}'\mu + 3\mathcal{W}\mu' + \frac{1}{12}\qty(8\mathcal{M}\mathcal{M}' - 2\mathcal{M}''')\varrho + \frac{1}{12}\qty(8\mathcal{M}^2-9\mathcal{M}'')\varrho' -\frac{5}{4}\mathcal{M}'\varrho'' - \frac{5}{6}\mathcal{M}\varrho'''+ \frac{1}{6}\varrho^{(5)}\,,\notag\\
    %-------------------------------------------------------------------------
    \dot{\mathcal{V} }&=\mathcal{W'}\xi+3\mathcal{W}\xi' +3\mathcal{V}\mu' +\frac{1}{12}(8\mathcal{NM'}+8\mathcal{MN'}-2\mathcal{N}''')\varrho+\frac{1}{12}(16\mathcal{MN}-9\mathcal{N''})\varrho'\notag\\
    &-\frac{5}{4}\mathcal{N}'\varrho''-\frac{5}{6}\mathcal{N}\varrho'''+\frac{1}{12}(8\mathcal{MM'}-2\mathcal{M}''')\varepsilon+\frac{1}{12}(8\mathcal{M}^2-9\mathcal{M}'')\varepsilon'-\frac{5}{4}\mathcal{M}'\varepsilon''-\frac{5}{6}\mathcal{M}\varepsilon'''+ \frac{1}{6}\varepsilon^{(5)}\notag\\
    %-------------------------------------------------------------------------
    \dot{\mathcal{X} }&=\mathcal{W'}\vartheta+3\mathcal{W}\vartheta'+\mathcal{V}'\xi+3\mathcal{V}\xi'+\mathcal{X'}\mu+3\mathcal{X}\mu'+\frac{1}{12}\qty(8\mathcal{MF'}+8\mathcal{FM'}+8\mathcal{NN'}-2\mathcal{F}''')\varrho\notag\\
    &+\frac{1}{12}\qty(16\mathcal{FM}+8\mathcal{N}^2-9\mathcal{F}'')\varrho'-\frac{5}{4}\mathcal{F'}\varrho''-\frac{5}{6}\mathcal{F}\varrho'''+\frac{1}{12}\qty(8\mathcal{NM'}+8\mathcal{MN'}-2\mathcal{N}''')\varepsilon\notag\\
    &+\frac{1}{12}\qty(16\mathcal{MN}-9\mathcal{N}'')\varepsilon'-\frac{5}{4}\mathcal{N}'\varepsilon''-\frac{5}{6}\mathcal{N}\varepsilon'''\notag\\
    &+\frac{1}{12}\qty(8\mathcal{MM'}-2\mathcal{M}''')\varphi+\frac{1}{12}\qty(8\mathcal{M}^2-9\mathcal{M}'')\varphi'-\frac{5}{4}\mathcal{M}'\varphi''-\frac{5}{6}\mathcal{M}\varphi'''+\frac{1}{6}\varphi^{(5)}\,,  
 \label{translawu}
\end{align}
where dot corresponds to the derivative with respect to $u$. 

In summary, the asymptotic behaviour is described by gauge fields of the form given in \eqref{AsMF}, where the components $a_{\phi}$ and $a_u$ of the asymptotic gauge field $a$ are given by \eqref{aa} and \eqref{AsMu}, respectively.

The structure outlined above provides insights into the asymptotic symmetries of the higher spin Maxwell Chern-Simons gravity and its associated algebra. In fact, the charge algebra of this theory can be derived using the Regge-Teitelboim approach \cite{Regge:1974zd}. In the following sections, we will explore this construction in greater detail.

\subsection{Charge algebra: spin-3 extension of the $\mathfrak{max}$-$\mathfrak{bms}_{3}$ algebra}
The charge algebra of the three-dimensional HS Maxwell CS gravity theory in representation of Poisson brackets can be obtained using the Regge-Teitelboim method \cite{Regge:1974zd} directly from the transformation law
\begin{equation}\label{rt}
\delta_{\Lambda_{2}}Q[\Lambda_{1}]=\left\{ Q[\Lambda_{1}],Q[\Lambda_{2}]\right\} ,
\end{equation}
where $Q[\Lambda]$ are the conserved charges spanning the algebra \cite{Banados:1994tn}. On the other hand, the variation of the charge in a CS theory is given by 
\begin{equation}
\delta Q[\Lambda]=\frac{k}{2\pi}\int\limits _{\partial\Sigma}\left\langle \Lambda\delta A\right\rangle \,.
\end{equation}
After applying the gauge transformation \eqref{AsMF} which introduces the asymptotic gauge field (\ref{aa}) we get \cite{Banados:1998gg}
\begin{equation}
\delta Q[\lambda]=\frac{k}{2\pi}\int d\phi\left\langle \lambda\delta a_{\phi}\right\rangle \,,\label{QQ}
\end{equation}
where $\Lambda=h^{-1}\lambda h$. Considering the invariant tensor \eqref{it1}-\eqref{it2} and the gauge field $a_\phi$ defined in \eqref{aa} in the previous expression, we get
\begin{equation}\label{charge-variation-01}
    \delta Q [y,f,h,v,w,g ]= \int d\phi \left( y\delta \mathbf{J}+ f\delta \mathbf{P} + h\delta \mathbf{Z}+v\delta \mathbf{V}+w\delta \mathbf{W}+g\delta \mathbf{X} \right),
\end{equation}
where we have defined
\begin{align}
   \mathbf{J} &= \frac{k}{4\pi}\left(\alpha_0\mathcal{M}+\alpha_1\mathcal{N}+\alpha_2\mathcal{F}\right)\, , \notag\\
   \mathbf{P}&=  \frac{k}{4\pi}\left(\alpha_{1}\mathcal{M}+ \alpha_{2}\mathcal{N}\right)\,,  \notag\\
   \mathbf{Z}&= \frac{k}{4\pi}\alpha_{2} \mathcal{M}\,,  \notag\\
\mathbf{V} &= \frac{k}{\pi}\left(\alpha_0\mathcal{W}+\alpha_1\mathcal{V}+\alpha_2\mathcal{X}\right)\, , \notag\\
\mathbf{W}&=  \frac{k}{\pi}\left(\alpha_{1}\mathcal{W}+ \alpha_{2}\mathcal{V}\right)\,, \notag\\
 \mathbf{X}&= \frac{k}{\pi}\alpha_{2} \mathcal{W}\,.
\end{align}
Assuming that the functions $y$, $f$, $h$, $v$, $w$ and $g$ do not depend on the fields we can directly integrate the variation on the phase space, and we find
\begin{align}
 Q [y,f,h,v,w,g ]= \int d\phi \left( y\mathbf{J}+ f \mathbf{P} + h \mathbf{Z}+v \mathbf{V}+w \mathbf{W}+g \mathbf{X} \right).\label{QQ2}
\end{align}
We can see that there are six independent surface charges,
% \begin{align}
%     j[y] &= Q[y,0,0,0,0,0] \, , & p[f] &= Q[0,f,0,0,0,0] \, , & z[h] &= Q[0,0,h,0,0,0] \,, \nonumber \\
%    V[v]&=Q[0,0,0,v,0,0]\,, & W[w] &= Q[0,0,0,0,w,0] \, , & X[g] &= Q[0,0,0,0,0,g]
% \end{align}
associated with six independent symmetry generators $y$, $f$, $h$, $v$, $w$ and $g$. Then, the Poisson brackets of these independent charges can be evaluated considering
\begin{equation}\label{rt2}
\delta_{\lambda_{2}}Q[\lambda_{1}]=\left\{ Q[\lambda_{1}],Q[\lambda_{2}]\right\} \,.
\end{equation}
% and \eqref{translaw}. 
As it is expected, the asymptotic symmetries associated to $y$, $f$ and $h$ span the $\mathfrak{max}$-$\mathfrak{bms}_3$ algebra \cite{Concha:2018zeb}. Indeed, expanding in Fourier modes according to
 \begin{align}
X=\frac{1}{2\pi}\sum X_m e^{im \phi}\,,
 \end{align}
one obtain the following algebra:
\begin{align}
 i\left\{ J_{m},J_{n}\right\} & =  \left(m-n\right)J_{m+n}+ \alpha_0\,k\,m^{3}\delta_{m+n,0}\,,\nonumber\\
i\left\{ J_{m},P_{n}\right\}  & =  \left(m-n\right)P_{m+n}+\alpha_1\,k\,m^{3}\delta_{m+n,0}\,,\nonumber\\
i\left\{ {P}_{m},P_{n}\right\}  & =  \left(m-n\right){Z}_{m+n}+\alpha_2\,k\,m^{3}\delta_{m+n,0}\,,\nonumber\\
i\left\{ {J}_{m},{Z}_{n}\right\}  & =  \left(m-n\right){Z}_{m+n}+\alpha_2\,k\,m^{3}\delta_{m+n,0}\,, \nonumber\\
i\left\{ {P}_{m},{Z}_{n}\right\}  & = 0\,, \nonumber \\
i\left\{ {Z}_{m},{Z}_{n}\right\}  & = 0\,,
\label{asymhsmax1}
\end{align}
which corresponds to the asymptotic symmetry algebra for the three-dimensional Maxwell CS gravity derived in \cite{Caroca:2017onr} and subsequently in \cite{Concha:2018zeb}.
The brackets of the previous Fourier modes with the other charges associated to $v$, $w$ and $g$ read
\begin{align}
 i\left\{ J_{m},V_{n}\right\} & =  \left(2m-n\right)V_{m+n}\,, &
i\left\{ J_{m},W_{n}\right\}  & =  \left(2m-n\right)W_{m+n}\,,\nonumber\\
i\left\{ P_{m},V_{n}\right\}  & =  \left(2m-n\right)W_{m+n}\,, &
i\left\{ {P}_{m},W_{n}\right\}  & =\left(2m-n\right)X_{m+n}\,,\nonumber\\
i\left\{ {J}_{m},{X}_{n}\right\}  & =  \left(2m-n\right){X}_{m+n}\,, &
i\left\{ {Z}_{m},{V}_{n}\right\}  & =  \left(2m-n\right){X}_{m+n}\,,\notag \\
i\left\{ {P}_{m},
X_{n}\right\}  & =0\,, \notag\\
i\left\{ {Z}_{m},
W_{n}\right\}  & =0\,, \notag\\
i\left\{ {Z}_{m},
X_{n}\right\}  & =0\,,
\label{asymhsmax2}
\end{align}
where we have defined the Fourier modes:
\begin{align}
{V}_{m}&=Q[v=e^{im\phi}]\,, & {W}_{m}&=Q[w=e^{im\phi}]\,,& {X}_{m}=Q[g=e^{im\phi}]\,,
\end{align}
which correspond to spin-3 generators.
Finally, these generators satisfy the following brackets:
\begin{align}
i\left\{ W_{m},W_{n}\right\}  & =\frac{8}{3\alpha_2 k}(m-n)\Omega_{m+n}^{ZZ}+\frac{1}{3}(m-n)(2m^{2}+2n^2-nm)Z_{m+n}+\frac{\alpha_2 k}{3}\,m ^{5}\delta_{m+n,0}\,,\nonumber\\
i\left\{ W_{m},V_{n}\right\}  & =  \frac{8}{3\alpha_2 k}(m-n)\left(2\Omega_{m+n}^{PZ}-\frac{\alpha_1}{\alpha_2}\Omega_{m+n}^{ZZ}\right)+\frac{1}{3}(m-n)(2m^{2}+2n^2-nm)P_{m+n}+\frac{\alpha_1 k}{3}\,m ^{5}\delta_{m+n,0}\,,\nonumber\\
i\left\{ V_{m},X_{n}\right\}  & =\frac{8}{3\alpha_2 k}(m-n)\Omega_{m+n}^{ZZ}+\frac{1}{3}(m-n)(2m^{2}+2n^2-nm)Z_{m+n}+\frac{\alpha_2 k}{3}\,m ^{5}\delta_{m+n,0}]\,,\nonumber\\
i\left\{ V_{m},V_{n}\right\} & =\frac{8}{3\alpha_2 k}(m-n)\left[\frac{1}{\alpha_2}\left(\frac{\alpha_1^2}{\alpha_2}-\alpha_0\right)\Omega_{m+n}^{ZZ}+\Omega_{m+n}^{PP}-2\frac{\alpha_1}{\alpha_2}\Omega_{m+n}^{PZ}+2\Omega_{m+n}^{ZJ}\right] \notag\\
&+\frac{1}{3}(m-n)(2m^{2}+2n^2-nm)J_{m+n}+ \frac{\alpha_0 k}{3}\,m ^{5}\delta_{m+n,0}\,,\nonumber\\
i\left\{ {W}_{m},{X}_{n}\right\}  & = 0\,,\notag \\
i\left\{ {X}_{m},{X}_{n}\right\}  & =  0\,,
\label{asymhsmax3}
\end{align}
where we have defined the following terms
\begin{equation}
    \Omega_{m}^{T\Bar{T}}=\sum_{j=-\infty}^{\infty}T_j \Bar{T}_{m-j}\,.
\end{equation}
The previous algebra corresponds to a higher spin-extension of the $\mathfrak{max}$-$\mathfrak{bms}_3$ found in \cite{Concha:2018zeb}, which we denote as $\mathfrak{hs}_3\mathfrak{max}$-$\mathfrak{bms}_3$. Note that the nonlinearity of the algebra is not trivial and does not simply correspond to a Maxwell generalization of the $\mathfrak{bms}_3$ algebra. Furthermore, in this case we have three central terms switched on, as can be seen by defining 
\begin{equation}
    c_i=12k\alpha_{i-1}\,, \qquad \text{with}\, \,  i=1,2,3.
\end{equation}
With this definition of the central charges, the algebra can be written as
\begin{align}
 i\left\{ J_{m},J_{n}\right\} & =  \left(m-n\right)J_{m+n}+ \frac{c_1}{12}\,m^{3}\delta_{m+n,0}\,,\nonumber\\
i\left\{ J_{m},P_{n}\right\}  & =  \left(m-n\right)P_{m+n}+\frac{c_2}{12}\,m^{3}\delta_{m+n,0}\,,\nonumber\\
i\left\{ {P}_{m},P_{n}\right\}  & =  \left(m-n\right){Z}_{m+n}+\frac{c_3}{12}\,m^{3}\delta_{m+n,0}\,,\nonumber\\
i\left\{ {J}_{m},{Z}_{n}\right\}  & =  \left(m-n\right){Z}_{m+n}+\frac{c_3}{12}\,m^{3}\delta_{m+n,0}\,,
\label{asymhsmax11}
\end{align}
\begin{align}
 i\left\{ J_{m},V_{n}\right\} & =  \left(2m-n\right)V_{m+n}\,, &
i\left\{ J_{m},W_{n}\right\}  & =  \left(2m-n\right)W_{m+n}\,,\nonumber\\
i\left\{ P_{m},V_{n}\right\}  & =  \left(2m-n\right)W_{m+n}\,, &
i\left\{ {P}_{m},W_{n}\right\}  & =\left(2m-n\right)X_{m+n}\,,\nonumber\\
i\left\{ {J}_{m},{X}_{n}\right\}  & =  \left(2m-n\right){X}_{m+n}\,, &
i\left\{ {Z}_{m},{V}_{n}\right\}  & =  \left(2m-n\right){X}_{m+n}\,,
\label{asymhsmax22}
\end{align}

\begin{align}
i\left\{ W_{m},W_{n}\right\}  & =\frac{32}{c_3}(m-n)\Omega_{m+n}^{ZZ}+\frac{1}{3}(m-n)(2m^{2}+2n^2-nm)Z_{m+n}+\frac{c_3}{36}\,m ^{5}\delta_{m+n,0}\,,\nonumber\\
i\left\{ W_{m},V_{n}\right\}  & = \frac{32}{c_3}(m-n)\left(2\Omega_{m+n}^{PZ}-\frac{c_2}{c_3}\Omega_{m+n}^{ZZ}\right)+\frac{1}{3}(m-n)(2m^{2}+2n^2-nm)P_{m+n}+\frac{c_2}{36}\,m ^{5}\delta_{m+n,0}\,,\nonumber\\
i\left\{ V_{m},X_{n}\right\}  & =\frac{32}{c_3}(m-n)\Omega_{m+n}^{ZZ}+\frac{1}{3}(m-n)(2m^{2}+2n^2-nm)Z_{m+n}+\frac{c_3}{36}\,m ^{5}\delta_{m+n,0}]\,,\nonumber\\
i\left\{ V_{m},V_{n}\right\} & =\frac{32}{c_3}(m-n)\left[\frac{1}{c_3}\left(\frac{c_2^2}{c_3}-c_1\right)\Omega_{m+n}^{ZZ}+\Omega_{m+n}^{PP}-2\frac{c_2}{c_3}\Omega_{m+n}^{PZ}+2\Omega_{m+n}^{ZJ}\right] \notag\\
&+\frac{1}{3}(m-n)(2m^{2}+2n^2-nm)J_{m+n}+ \frac{c_1}{36}\,m ^{5}\delta_{m+n,0}\,.
\label{asymhsmax33}
\end{align}
In sum, the commutation relations \eqref{asymhsmax11}-\eqref{asymhsmax33} provides the asymptotic symmetries of spin-3 fields coupled to Maxwell CS gravity. In the next section, we show that this algebra can alternatively be recovered as the flat limit of three copies of the $\mathcal{W}_3$ algebra, mimicking the relation between HS Maxwell gravity and three copies of $\mathfrak{sl}(3,\mathbb{R})$.

\section{Spin-3 extension of the $\mathfrak{max}$-$\mathfrak{bms}_3$ algebra as a vanishing cosmological constant limit}

It is possible to check that the asymptotic symmetries described by three copies of the  $\mathcal{W}_3$ algebra lead to the spin-3 extension of the $\mathfrak{max}$-$\mathfrak{bms}_3$ algebra previously obtained. Indeed, three copies of the $\mathcal{W}_3$ algebra can be written as
\begin{align}
i\left\{ \mathcal{L}_{m}^{\pm},\mathcal{L}_{n}^{\pm}\right\}  & =  \left(
m-n\right) \mathcal{L}_{m+n}^{\pm}+\dfrac{c^{\pm}}{12}\,m^3
\delta _{m+n,0}\,, \notag\\
i\left\{ \hat{\mathcal{L}}_{m},\hat{\mathcal{L}}_{n}\right\}  & =  \left(
m-n\right) \mathcal{\hat{L}}_{m+n}+\dfrac{\hat{c}}{12}\,m^3\delta _{m+n,0}\,, \notag\\
i\left\{ \mathcal{L}_{m}^{\pm},\mathcal{W}_{n}^{\pm}\right\}  & = \left(
2m-n\right) \mathcal{W}_{m+n}^{\pm}\,, \notag\\
i\left\{ \hat{\mathcal{L}}_{m},\hat{\mathcal{W}}_{n}\right\}  & = \left(
2m-n\right) \hat{\mathcal{W}}_{m+n}\,, \notag\\
i\left\{ \mathcal{W}_{m}^{\pm},\mathcal{W}_{n}^{\pm}\right\}  & = \frac{1}{3}\left[\frac{96}{c^{\pm}}(m-n)\Omega_{m+n}^{\pm}+(m-n)(2m^2-2n^2-mn)\mathcal{L}_{m+n}^{\pm}+\frac{c^{\pm}}{12}m^5\delta_{m+n,0}\right]\,,\notag \\
i\left\{ \hat{\mathcal{W}}_{m},\hat{\mathcal{W}}_{n}\right\}  & = \frac{1}{3}\left[\frac{96}{\hat{c}}(m-n)\hat{\Omega}_{m+n}+(m-n)(2m^2-2n^2-mn)\hat{\mathcal{L}}_{m+n}+\frac{\hat{c}}{12}m^5\delta_{m+n,0}\right]\,,
\end{align}
with
\begin{equation}
\Omega_{m}^{\pm}\equiv\sum_{j\in\mathbb{Z}}\mathcal{L}_{m+j}^{\pm}\mathcal{L}_{-j}^{\pm}\,, \qquad \hat{\Omega}_{m}\equiv\sum_{j\in\mathbb{Z}}\hat{\mathcal{L}}_{m+j}\hat{\mathcal{L}}_{-j}\,.
\end{equation}
Then, by redefining the generators as
\begin{align}
   J_{m}&=\mathcal{L}_{m}^{+}-\mathcal{L}_{-m}^{-}-\hat{\mathcal{L}}_{-m}\,,&
    P_{m}&=\frac{1}{\ell}\left(\mathcal{L}_{m}^{+}+\mathcal{L}_{-m}^{-}\right)\,,&
    Z_{m}&=\frac{1}{\ell^2}\left(\mathcal{L}_{m}^{+}-\mathcal{L}_{-m}^{-}\right)\,,\notag\\
    V_{m}&=\mathcal{W}_{m}^{+}-\mathcal{W}_{-m}^{-}-\hat{\mathcal{W}}_{-m}\,,&
    W_{m}&=\frac{1}{\ell}\left(\mathcal{W}_{m}^{+}+\mathcal{W}_{-m}^{-}\right)\,,&
    X_{m}&=\frac{1}{\ell^2}\left(\mathcal{W}_{m}^{+}-\mathcal{W}_{-m}^{-}\right)\,,\label{cob}        
\end{align}
along the redefinition of the central charges as follows,
\begin{align}
c_1&=c^+-c^--\hat{c},\, & c_2&=\frac{1}{\ell}(c^++c^-),\, & c_3=\frac{1}{\ell^2}(c^+-c^-),\,
\end{align}
the centrally extended $\mathcal{W}_3\oplus\mathcal{W}_3\oplus\mathcal{W}_3$ can be written as
\begin{align}
 i\left\{ J_{m},J_{n}\right\} & =  \left(m-n\right)J_{m+n}+\dfrac{c_{1}}{12}\,m^{3}\delta_{m+n,0}\,,\nonumber\\
i\left\{ J_{m},P_{n}\right\}  & =  \left(m-n\right)P_{m+n}+\dfrac{c_{2}}{12}\,m^{3}\delta_{m+n,0}\,,\nonumber\\
i\left\{ P_{m},P_{n}\right\}  & =  \left(m-n\right)Z_{m+n}+\dfrac{c_{3}}{12}\,m^{3}\delta_{m+n,0}\,,\nonumber\\
i\left\{ J_{m},Z_{n}\right\}  & =  \left(m-n\right)Z_{m+n}+\dfrac{c_{3}}{12}\,m^{3}\delta_{m+n,0}\,,\nonumber\\
i\left\{ P_{m},Z_{n}\right\} & =  \frac{1}{\ell^{2}}\left(m-n\right)P_{m+n}+\dfrac{c_{2}}{12\ell^{2}}\,m^{3}\delta_{m+n,0}\,,\nonumber\\
i\left\{ Z_{m},Z_{n}\right\} & =  \frac{1}{\ell^{2}}\left(m-n\right)Z_{m+n}+\dfrac{c_{3}}{12\ell^{2}}\,m^{3}\delta_{m+n,0}\,,\label{asymhsadsL1}
\end{align}
\begin{align}
 i\left\{ J_{m},V_{n}\right\} & =  \left(2m-n\right)V_{m+n}\,, &
i\left\{ J_{m},W_{n}\right\}  & =  \left(2m-n\right)W_{m+n}\,,\nonumber\\
i\left\{ P_{m},V_{n}\right\}  & =  \left(2m-n\right)W_{m+n}\,, &
i\left\{ {P}_{m},W_{n}\right\}  & =\left(2m-n\right)X_{m+n}\,,\nonumber\\
i\left\{ {J}_{m},{X}_{n}\right\}  & =  \left(2m-n\right){X}_{m+n}\,, &
i\left\{ {Z}_{m},{V}_{n}\right\}  & =  \left(2m-n\right){X}_{m+n}\,,\notag \\
i\left\{ {P}_{m},
X_{n}\right\}  & =\frac{1}{\ell^2}\left(2m-n\right)W_{m+n}\,, & 
i\left\{ {Z}_{m},
W_{n}\right\}  & =\frac{1}{\ell^2}\left(2m-n\right){W}_{m+n}\,, \notag\\
i\left\{ {Z}_{m},
X_{n}\right\}  & =\frac{1}{\ell^2}\left(2m-n\right){X}_{m+n}\,,
\label{asymhsadsL2}
\end{align}
\begin{align}
i\left\{ W_{m},W_{n}\right\}  & =\frac{32}{(c_2^2/\ell^2-c_3^2)}(m-n)\left[2\frac{c_2}{\ell^2}\Omega_{m+n}^{PZ}-c_3\left(\Omega_{m+n}^{ZZ}+\frac{\Omega_{m+n}^{PP}}{\ell^2}\right)\right]\notag \\
&+\frac{1}{3}(m-n)(2m^{2}+2n^2-nm)Z_{m+n}
+\frac{c_3}{36}\,m ^{5}\delta_{m+n,0}\,,\nonumber\\
i\left\{ W_{m},V_{n}\right\}  & = \frac{32}{c_2^2/\ell^2-c_3^2}(m-n)\left[-2c_3\Omega_{m+n}^{PZ}+c_2\left(\Omega_{m+n}^{ZZ}+\frac{\Omega_{m+n}^{PP}}{\ell^2}\right)\right] \notag \\
&+\frac{1}{3}(m-n)(2m^{2}+2n^2-nm)P_{m+n}
+\frac{c_2}{36}\,m ^{5}\delta_{m+n,0}\,,\nonumber\\
i\left\{ V_{m},X_{n}\right\}   & =\frac{32}{(c_2^2/\ell^2-c_3^2)}(m-n)\left[2\frac{c_2}{\ell^2}\Omega_{m+n}^{PZ}-c_3\left(\Omega_{m+n}^{ZZ}+\frac{\Omega_{m+n}^{PP}}{\ell^2}\right)\right]\notag \\
&+\frac{1}{3}(m-n)(2m^{2}+2n^2-nm)Z_{m+n}
+\frac{c_3}{36}\,m ^{5}\delta_{m+n,0}\,,\nonumber\\
i\left\{ W_{m},X_{n}\right\}   & =\frac{32}{(c_2^2/\ell^2-c_3^2)\ell^2}(m-n)\left[-2c_3\Omega_{m+n}^{PZ}+c_2\left(\Omega_{m+n}^{ZZ}+\frac{\Omega_{m+n}^{PP}}{\ell^2}\right)\right]\notag \\
&+\frac{1}{3\ell^2}(m-n)(2m^{2}+2n^2-nm)P_{m+n}
+\frac{c_2}{36\ell^2}\,m ^{5}\delta_{m+n,0}\,,\nonumber\\
i\left\{ X_{m},X_{n}\right\}  & =\frac{32}{(c_2^2/\ell^2-c_3^2)\ell^2}(m-n)\left[2\frac{c_2}{\ell^2}\Omega_{m+n}^{PZ}-c_3\left(\Omega_{m+n}^{ZZ}+\frac{\Omega_{m+n}^{PP}}{\ell^2}\right)\right]\notag \\
&+\frac{1}{3\ell^2}(m-n)(2m^{2}+2n^2-nm)Z_{m+n}
+\frac{c_3}{36\ell^2}\,m ^{5}\delta_{m+n,0}\,,\nonumber\\
i\left\{ V_{m},V_{n}\right\}  & =\frac{32}{(-c_1c_2^2/\ell^4+c_2c_3/\ell^2+c_1c_3^2/\ell^2-c_3^3)}(m-n)\left[-2\frac{c_1c_2}{\ell^2}\Omega_{m+n}^{PZ}\right.\notag\\
&\left.+c_3^2\left(-\Omega_{m+n}^{PP}+\frac{\Omega_{m+n}^{JJ}}{\ell^2}-2\Omega_{m+n}^{JZ}\right)+c_1c_3\left(\Omega_{m+n}^{ZZ}+\frac{\Omega_{m+n}^{PP}}{\ell^2}\right)\right.\notag \\
&\left.-c_2^2\left(\frac{\Omega_{m+n}^{JJ}}{\ell^4}-2\frac{\Omega_{m+n}^{JZ}}{\ell^2}+\Omega_{m+n}^{ZZ}\right)\right]\notag \\
&+\frac{1}{3}(m-n)(2m^{2}+2n^2-nm)J_{m+n}
+\frac{c_1}{36}\,m ^{5}\delta_{m+n,0}\,.
\label{asymhsadsL3}
\end{align}
Then, taking the flat limit $\ell\rightarrow \infty$, the asymptotic symmetry algebra given by three copies of the $\mathcal{W}_3$ algebra leads to the $\mathfrak{hs}_3\mathfrak{max}$-$\mathfrak{bms}_3$ obtained in the previous section. The algebra \eqref{asymhsadsL1}-\eqref{asymhsadsL3} is a spin-3 extension of the enlarged $\mathfrak{bms}_3$ algebra introduced in \cite{Concha:2018jjj}, and corresponds to the asymptotic symmetry algebra of AdS-Lorentz CS gravity coupled to a spin$-3$ field \cite{concha:2025}.

%%%%%%%%%%%%%%%%%%%%%%%%%%%%%%%%
\section{Discussion}\label{concl}

In this paper, we have studied the asymptotic structure of the Chern-Simons gravity theory invariant under a higher spin extension of the so-called Maxwell symmetry. We proposed some consistent asymptotic conditions that allow to canonically realized a nonlinear higher-spin extension of the Maxwell generalization of the $\mathfrak{bms}_{3}$ symmetry that we call $\mathfrak{hs}_3\mathfrak{max}$-$\mathfrak{bms}_{3}$. It was also shown that this algebra can be obtained as a vanishing cosmological constant limit of three copies of the $\mathcal{W}_3$ algebra, with three independent central charges.

One of the most fascinating aspects of three-dimensional higher-spin gravities is the presence of exact solutions that carry higher-spin charges. This work serves as a foundational step in exploring the thermodynamics of such configurations through topological arguments \cite{Crisostomo:2000bb,Matulich:2014hea}. In the present case we expect to find a novel higher-spin extension of locally flat cosmological spacetimes. Since we have incorporated the chemical potential conjugated to the higher spin charges, the thermodynamics properties of the solution can be analyzed along the lines of \cite{Matulich:2014hea}. An interesting direction now available for exploration involves the potential application of the description of the theory in terms of three copies of $SL(3,R)$. This framework may provide insights in the context of thermodynamic properties of either black hole solutions of (higher-spin extension of) AdS-Lorentz gravity or cosmological solutions of (higher-spin extension of) Maxwell gravity.  

A possible extension of our results can be carried out by incorporating a negative cosmological constant to the higher-spin Maxwell CS theory. This can be done by considering the higher-spin extension of the so-called AdS-Lorentz gravity constructed in \cite{Caroca:2017izc}. In this case, we expect to find higher-spin black holes generalizing the BTZ-type solution of the AdS-Lorentz gravity studied in \cite{Concha:2018jjj} including spin-3 charges.  Then, the higher spin black hole solution should lead to the higher spin extension of locally flat cosmological spacetimes, after performing the vanishing cosmological constant limit. In this scenario, we could also study the solutions and thermodynamics of the higher-spin extension of the Maxwell teleparallel gravity first presented in \cite{Adami:2020xkm} and analyse how the presence of a non-vanishing torsion modifies the solutions of the spin-3 extension of the Maxwell gravity theory. As solutions we expect to find a spin-3 extension of a Maxwell teleparallel black hole [work in progress]. We guess that both spin-3 black holes will reduce to the Maxwellian generalization of the spin-3 extension of flat cosmologies in a flat limit.

Another interesting aspect that it would be worthwhile to explore is the study of the asymptotic symmetry of the three-dimensional Maxwell CS gravity theory coupled to spin-$5/2$ gauge field \cite{Caroca:2021bjo,Caroca:2023oie}. In such case, the HS field theory contains fields of spins 2, 4 and $5/2$ whose invariance is extended to HS fermionic symmetry transformations, referred as hypersymmetry \cite{Aragone:1983sz,Zinoviev:2014sza,Henneaux:2015tar,Henneaux:2015ywa,Fuentealba:2015jma,Fuentealba:2015wza}. Following the results obtained here, one could expect to find, after imposing suitable boundary conditions, a deformation of the hyper $\mathfrak{bms}_{3}$ which should be alternatively recovered as a vanishing cosmological constant limit of a precise combination of the $\mathcal{W}_{\left(2,\frac{5}{2},4\right)}$ and $\mathcal{W}_{\left(2,4\right)}$ algebra \cite{Bellucci:1994xa,Figueroa-OFarrill:1991huu}. The asymptotic algebra could serve to derive hypersymmetry bounds which could imply interesting properties for solutions as in \cite{Henneaux:2015tar,Fuentealba:2015jma}.

% - future:
% - canonical realization of the full case before the limit. 
% - 

%%%%%%%%%%%%%%%%%%%%%%%%%%%%%%%%%%%%%%%%%%%%%%%%%%%%%%%%%%%%%%%%%%%%%%%%%%%%%%%%%%%%%%%%%%%%%

\section*{Acknowledgment}
This work was funded by the National Agency for Research and Development ANID - SIA grant No. SA77210097 and FONDECYT grant 11220328, 11220486, 1211077 and 1231133. PC and ER would also like to thank the Dirección de Investigación and Vice-rectoría de Investigación of the Universidad Católica de la Santísima Concepción, Chile, for their constant support. J.M. has been supported by the MCI, AEI, FEDER (UE)  grants PID2021-125700NB-C21 (“Gravity, Supergravity and Superstrings”(GRASS)) and IFT Centro de Excelencia Severo Ochoa CEX2020-001007-S. D.P. would like to thank to Universidad de Concepción, Chile, for Beca articulación pregrado-postgrado.
%\section*{Appendix} \label{app}
%%%%%%%%%%%%%%%%%%%%%%%%%%%%%%%%%%%%%%%%%%%%%%%%%%%%%%%%%%%%%%%%%%%%%%%%%%%%%%%%%%%%%%%%%%%%%%%%%%%%

\bibliographystyle{fullsort}
 
\bibliography{Maxwell_Spin3}

\end{document}